\documentclass[12pt]{article}

\usepackage{a4,graphics,graphicx,amsmath,amssymb,cite,color}
\usepackage[verbose]{wrapfig}
\catcode`@=11 \@addtoreset{equation}{section} \catcode`@=12

\begin{document}
\title{
{\baselineskip -.2in
\vbox{\small\hskip 4in \hbox{IITM/PH/TH/2011/4}}} 
\vskip .4in
\vbox{
{\bf \large On the Stability of Non-Supersymmetric Quantum Attractors in String Theory}
}
\author{Pramod Dominic${}$\thanks{email: pramod@physics.iitm.ac.in}
~and
Prasanta K. Tripathy${}$\thanks{email: prasanta@physics.iitm.ac.in} \\
\normalsize{\it ${}$ Department of Physics,}\\
\normalsize{\it Indian Institute of Technology Madras} \\
\normalsize{\it  Chennai 600 036, India.} 
}}
\maketitle

\begin{abstract}
We study four dimensional non-supersymmetric attractors in type $IIA$ string theory in the presence 
of sub-leading corrections to the prepotential. For a given Calabi-Yau manifold, the $D0-D4$ system
admits an attractor point in the moduli space which is  uniquely specified by the black hole charges.
The perturbative corrections to the prepotential do not change the number of massless directions in 
the black hole effective potential. We further study non-supersymmetric $D0-D6$ black holes in the 
presence of sub-leading corrections.  In this case the space of attractor points define a hypersurface 
in the moduli space. 
\end{abstract}
\renewcommand{\thefootnote}{\arabic{footnote}}
\setcounter{footnote}{0}
\newpage
\section{Introduction}
It is well known that string theory provides a microscopic understanding of the origin of black hole 
entropy\cite{Strominger:1996sh}. For a large class of intersecting brane configurations, it is possible 
to compute the leading
order contribution for the degeneracy in the limit when the string coupling is weak. They agree 
with the entropy of the corresponding supergravity  black holes, which exist in the appropriate 
weak curvature limits. Furthermore, there is a considerable progress in computing the sub-leading 
corrections to the microscopic counting of the degeneracy along with their macroscopic counterparts. 
For a recent review on these results see Ref.\cite{Sen:2007qy}.

The microscopic counting gives the degeneracy of states as a function of the quantized charges of 
the intersecting brane configurations. From the supergravity side, the fact that the entropy should 
depend only on the black hole charges, at least in the case of single centered black holes, is 
evident from the attractor mechanism. 

Soon after its revelation in the context of a simple supersymmetry preserving spherically symmetric 
black hole in  $N=2$ supergravity coupled to $n$ vector multiplets \cite{Ferrara:1995ih}, 
the  attractor mechanism has been studied extensively \cite{
Strominger:1996kf,Ferrara:1996dd,Ferrara:1996um,Ferrara:1997tw,Gibbons:1996af}.
The generalization of the attractor mechanism to supersymmetric, multi-centered black holes has 
been carried out  \cite{Denef:2000nb,Denef:2001xn,Denef:2002ru,Denef:2007vg}. It has also
been subsequently used to compute the higher derivative corrections to the black hole entropy 
\cite{Behrndt:1998eq,LopesCardoso:1998wt,LopesCardoso:1999cv,LopesCardoso:1999ur}.
A systematic procedure for computing the black hole entropy using attractor mechanism in higher 
derivative gravity has been developed  \cite{Sen:2005wa}.

Though the attractor mechanism was originally perceived by explicitly solving the spinor conditions, 
it was soon realized that the mechanism is in fact a consequence of the extremality of the black hole
\cite{Ferrara:1997tw,Gibbons:1996af}. 
This gave rise to the possibility of the existence of non-supersymmetric attractors along with their 
supersymmetric cousins. These non-supersymmetric attractors were explored in great detail in 
\cite{Goldstein:2005hq}. The existence of non-supersymmetric attractors in string theory was first 
shown in \cite{Tripathy:2005qp}. Various aspects of the non-supersymmetric attractors in string 
theory were subsequently studied \cite{Kallosh:2005ax,Kallosh:2006bt,Kallosh:2006ib,Goldstein:2005rr,Alishahiha:2006ke,Giryavets:2005nf,Astefanesei:2006dd,Dabholkar:2006tb,Saraikin:2007jc}.
Non-supersymmetric attractors were also shown to exist  in a large class of  gauged supergravity 
theories on symmetric spaces in four and five dimensions. New branches of solutions corresponding
to zero central charges were also explored in these models \cite{Andrianopoli:2006ub,Bellucci:2007ds,
Ferrara:2008hwa}. 
For some early results along these lines see 
Ref.\cite{Bellucci:2006ew,Ferrara:2006em,Ferrara:2006xx,Bellucci:2006xz,Ferrara:2006yb,D'Auria:2007ev,Andrianopoli:2007rm,Ferrara:2007pc,Ferrara:2007tu,Andrianopoli:2007kz}.
Derivation of non-supersymmetric attractors in certain class of models in terms of the 
extremization of a fake superpotential has been proposed \cite{Ceresole:2007wx,Andrianopoli:2007gt}. 
This prescription has helped in finding the full flow in a number of examples \cite{Lopes Cardoso:2007ky,Hotta:2007wz,Gaiotto:2007ag,Cai:1900ve,Goldstein:2008fq,Hotta:2009bm,Bossard:2009we,Galli:2009bj,Roychowdhury:2009ub,Mohaupt:2010fk,Nampuri:2010um,Siwach:2010zp,Marrani:2010bn,Galli:2010mg}. 
The generalization of the attractor mechanism to black holes in anti-deSitter as well as deSitter spaces 
has also been carried out \cite{Kimura:2008tq,Astefanesei:2007vh,Morales:2006gm,Goldstein:2009cv,Goldstein:2010aw,Dall'Agata:2010gj,Kachru:2011ps}.

As it has been emphasized in \cite{Tripathy:2005qp,Nampuri:2007gv}, the non-supersymmetric 
attractors differ from the supersymmetric ones in a crucial way.  The manifestation of the stability 
of the supersymmetric attractor is evident from the corresponding mass matrix of the effective black 
hole potential. On the other hand, for non-supersymmetric attractors $(n-1)$ of the $2 n$ real 
scalar fields remain massless. Thus, one might expect that the sub-leading corrections will drastically 
change the criterion for stability of such attractors. 

To understand the effect of the sub-leading corrections,  $N=2$ supergravity arising from the 
compactification of type $IIA$ string theory on a simple two-parameter Calabi-Yau manifolds  
was considered \cite{Dominic:2010yv}. It was noticed that, for non-supersymmetric $D0-D4$
black holes, the perturbative corrections to the prepotential do not lift the massless mode. It was
further suggested that in such cases, instanton corrections may lift the zero modes. 

The goal of the present work is to extent the above result on the perturbative sub-leading correction
to supergravity theories arising from string compactifications. 
We study both $D0-D4$ as well as $D0-D6$ configurations in presence of sub-leading corrections.
In both these cases, the perturbative quantum corrections do not change the number of zero
modes. Hence,
the role of non-perturbative corrections become crucial for both the systems.
The plan of the 
paper is as follows. In the next section we review some preliminaries on non-supersymmetric attractors 
in string theory  \cite{Tripathy:2005qp}. In \S3, we study the $D0-D4$ configuration. 
\S4 discusses the $D0-D6$ configuration in detail. We first review
the leading order results in \S4.1. In \S4.2 we  solve the equation of motion and in \S4.3 we derive 
the mass matrix and diagonalize. Finally, in \S5 we summarize the results  and in the appendix we 
workout some of the steps in detail.

\section{Attractors in type $IIA$ string theory}

In this section we will briefly review some of  the important results on   static, spherically symmetric 
non-supersymmetric  attractors in type $IIA$ string theory. Such solutions were first explored in
Ref. \cite{Tripathy:2005qp}. For the static, spherically symmetric black holes, the attractor point is 
obtained by extremizing  the effective black hole potential  \cite{Ferrara:1997tw}:
\begin{eqnarray}\label{veff}
V&=&e^K\left[g^{a\bar{b}}\nabla_aW\left(\nabla_bW\right)^*+|W|^2\right]
\end{eqnarray}
where $\nabla_aW=\partial_a W + \partial_aK W$ . 
Here $K$ is the K\"ahler potential in the moduli space. For $N=2$ theories, it is determined in 
terms of the prepotential $F$ as:
\begin{eqnarray}
K = - \ln{\Im}\left(\sum_{a=0}^N\bar{X}^{\bar a}\partial_a F(X)\right)
\label{kahler}
\end{eqnarray}
Here $\Im(f)$ denotes imaginary part of $f$. The superpotential $W$ for a configuration carrying $q_a$ electrically charged and $p^a$ 
magnetically charged branes is given by 
\begin{eqnarray}
W = \sum_{a=0}^N\Big(q_a X^a - p^a \partial_a F\Big) \ .
\label{suppot}
\end{eqnarray}
It is related to the central charge by $Z = e^{K/2} W$. Hence the supersymmetric attractor point is 
obtained  by solving the algebraic equation $\nabla_a W = 0$. The non-supersymmetric attractor
is obtained by extremizing  the black hole effective potential $\partial_a V = 0$ and finding its critical 
points for which $\nabla_a W \neq 0$.  For the potential (\ref{veff}), this condition becomes
\begin{equation}
\left(g^{b\bar c} \nabla_a\nabla_b W \overline{\nabla_cW} + 2 \nabla_a W \overline{W}
+ \partial_a g^{b\bar c} \nabla_b W\overline{\nabla_c W}\right) = 0~.
\label{critpt}
\end{equation}
The non-supersymmetric attractor becomes stable, if the quadratic term in the effective potential 
about the corresponding critical point becomes positive definite. Or, in other words, the $2n\times 2n$
matrix of  second derivatives of $V$ must admit only positive Eigenvalues.

We will be interested in $N=2$ supergravity theories obtained upon compactification of type $IIA$
string theory on a Calabi-Yau three-fold ${\cal M}$. In this case, the leading term of the prepotential, 
in the large volume limit is given by 
\begin{equation}
 F=D_{abc}\frac{X^aX^bX^c}{X^0} \ ,
 \label{leadingprep}
 \end{equation}
 where $D_{abc} = (1/3!) \int_{\cal M} J_a\wedge J_b\wedge J_c$ are the triple 
 intersection numbers associated with the Calabi-Yau manifold ${\cal M}$.  Here the two-forms 
 $\{J_a\}$ form a basis of the integral cohomology $H^2({\cal M},\mathbb{Z})$. The complex scalar 
 fields $X^a, a = 1,\cdots, n$, parametrize the vector multiplet moduli space.
 
For a general $D0-D4-D6$ configuration, the supersymmetric attractor for this case has been 
obtained  and it was shown that the macroscopic black hole entropy  computed using attractor 
mechanism agrees with the microscopic counting \cite{Maldacena:1997de}. In the case of 
non-supersymmetric attractor the solution was obtained in Ref. \cite{Tripathy:2005qp}. It was 
shown that the mass matrix at the non-supersymmetric critical point admits $(n+1)$ positive
and $(n-1)$ zero Eigenvalues. 

In view of the above, it is important to consider attractors in the presence of sub-leading corrections 
to the prepotential (\ref{leadingprep}). For a large number of two and three parameter models, 
holomorphic, perturbative corrections to the prepotential in  type $IIA$ supergravity has been 
computed using mirror symmetry  \cite{Hosono:1993qy,Hosono:1994ax,Hosono:1995bm}.  The 
prepotential has the general form:
 \begin{eqnarray}
 \label{fullprep}
F=D_{abc}\frac{X^aX^bX^c}{X^0}+ \alpha_{0a} X^a X^0 + \beta (X^0)^2 + {\rm nonperturbative \ terms} 
\end{eqnarray} 
with $\alpha_{0a} =  -(1/24)\int_{\cal M} c_2\wedge J_a$ and $\beta = -i {\zeta(3)}\chi/({16{\pi}^3})$. 
Here $c_2$ and $\chi$ are respectively the second Chern class and the Euler number of the Calabi-Yau 
three fold ${\cal M}$. The supersymmetric black holes in presences of these sub-leading terms were 
studied and the correction to the black hole entropy was obtained \cite{Behrndt:1996jn}. In the following 
sections we will discuss the effect of such corrections to the prepotential on the extremal non-supersymmetric black holes.

\section{The $D0-D4$ system}

In this section, we will consider the $D0-D4$ system in the presence of perturbative sub-leading 
corrections. To find the non-supersymmetric attractors we need to explicitly solve eq.(\ref{critpt}).
Let us first compute the K\"ahler potential (\ref{kahler}):
\begin{eqnarray}
K = - \ln{\Im}\left(\bar{X}^0\partial_0 F(X) + \sum_{a=1}^N\bar{X}^{\bar a}\partial_a F(X)\right) \ .
\nonumber
\end{eqnarray}
It is straightforward to see that the additional contributions from the sub-leading terms 
$\alpha_{0a} X^a X^0 + \beta (X^0)^2$ to $\bar{X}^0\partial_0 F(X) 
+ \sum_{a=1}^N\bar{X}^{\bar a}\partial_a F(X)$ is given by
$$ \alpha_{0a} (X^a \bar X^0 + \bar X^a X^0) + 2 \beta X^0\bar X^0 \ . $$
For Calabi-Yau manifolds which are obtained as complete intersections of product of projective spaces, 
the second Chern class is, $c_2 = c_2^{ab}\ J_a\wedge J_b$, where the  coefficients $c_2^{ab}$ are 
determined in terms of the degree and weights of various projective  coordinates \cite{Hosono:1994ax}. 
This implies that the  $\alpha_{0a}$s are all real and hence they do  not appear in the expression for the 
K\"ahler potential, which involves only the imaginary part of the  above quantity. On the other hand,
 $\beta$ is pure imaginary and hence can appear in the expression for the K\"ahler potential. However,
 as it was shown in Ref.\cite{Behrndt:1996jn}, it can be eliminated by a symplectic transformation. Hence
from now on we will ignore this term in the prepotential (\ref{fullprep}). 

Thus we find
 \begin{equation}
\label{IIAkp} 
K=-\ln\left(-iX^0{\bar X}^0 
D_{abc}
\left(\frac{X^a }{ X^0}-\frac{{\bar X}^a}{ {\bar X}^0}\right) 
\left(\frac{X^b} { X^0} -\frac{{\bar X}^b}{{\bar X}^0}\right) 
\left(\frac{X^c}{ X^0}-\frac{{\bar X}^c }{ {\bar X}^0}\right) 
\right) \ . \nonumber
\end{equation}
For the $D0-D4$ system, the superpotential $W$ is given by 
$$ W = q_0 X^0 - \sum_{a=1}^N p^a\partial_a F \ $$
Here $q_0$ is the charge of the $D0$ brane where as $p^a$ are the charges of $D4$ branes wrapped 
on the $4$-cycles of $\cal M$. After substituting for $\partial_a F$ from eq.(\ref{fullprep}), we find 
$$ W = q_0 X^0 - 3 D_{ab} X^a X^b/X^0 - \alpha_0 X^0 \ . $$
In the above we have introduced $D_{ab} = D_{abc} p^c$ and $\alpha_0 = \alpha_{0a} p^a$. We also
define $D_a = D_{ab} p^b$ and $D = D_a p^a$ for later use. 

For convenience, from now on we will introduce the scalar fields $x^a = X^a/X^0$ and set the gauge 
$X^0=1$.  The expression for the K\"ahler potential  $K$ and  superpotential $W$ in this gauge is 
given by:
\begin{eqnarray}
K & =& - \ln\left(-i D_{abc} (x^a - \bar x^a) (x^b - \bar x^b) (x^c - \bar x^c) \right) \ , \\
W & = & (q_0 - \alpha_0) - 3 D_{ab} x^a x^b \ .
\end{eqnarray}

We notice that the only change in the superpotential is a shift of the $D0$ brane charge by $\alpha_0$. 
Thus the computations will be identical to the one with the classical black hole with prepotential 
$F = D_{abc} X^a X^b X^c/X^0$. In particular, if we set the ansatz $x^a = p^a t$ we find that, for
$t\neq 0$, the equations of motion (\ref{critpt}) reduces to:
\begin{equation}
(q_0-\alpha_0 - t^2 D) (q_0 - \alpha_0 + t^2 D) = 0 \ . 
\end{equation}
Clearly, the supersymmetric solution corresponds to $t  = i  \sqrt{(q_0-\alpha_0)/D}$ where as the 
non-supersymmetric solution is given by $t  = i  \sqrt{(\alpha_0-q_0)/D}$. The supersymmetric solution
exists when $(q_0-\alpha_0)D>0$ where as the non-supersymmetric solution exists in the opposite 
domain $(\alpha_0-q_0)D>0$. The entropy of the black hole, in both the cases is given by  
$S = 2\pi \sqrt{|(q_0-\alpha_0)D|}$.

To understand the stability of the non-supersymmetric attractor, we need to consider fluctuations of 
the field $x^a$ around the attractor point: 
$$ x^a = i p^a  \sqrt{\frac{(\alpha_0-q_0)}{D}} + \delta\xi^a + i \delta y^a$$ and keep terms up to 
quadratic order in the black hole effective potential (\ref{veff}). We can read the mass matrix $M$ 
from \cite{Tripathy:2005qp}:
\begin{equation}
M =   48 e^{K_0} \frac{(\alpha_0-q_0)}{D}\left\{
\left(3 D_a D_b -  D  D_{ab} \right)\otimes I + D D_{ab}\otimes \sigma^3 \right\} \ .
\end{equation}
Here $I$ is the $2\times 2$ identity matrix, $\sigma^i$ are the Pauli matrices in the basis in which 
$\sigma^3$ is diagonal and  $K_0$ is the K\"ahler potential evaluated at the non-supersymmetric 
extremum. The only change from the leading result is in the pre-factor $e^{K_0}(-q_0/D)$ where 
$q_0$ is replaced by $(q_0-\alpha_0)$. Though there is a shift in mass for all the massive modes, 
the zero modes remain unchanged.  Thus, for $D(\alpha_0-q_0)>0$, the mass matrix still has 
$(n+1)$ positive Eigenvalues and $(n-1)$ zero Eigenvalues. 

In this section we have noticed that the non-supersymmetric attractor for $D0-D4$ system in the 
presence of perturbative sub-leading correction can be obtained from the leading order solution
quite trivially by a shift of the $D0$ brane charge. Hence all the results pertaining to the leading 
solution hold. Especially the condition for stability of the attractor does not change. In the next 
section we will study the $D0-D6$ system, where we will see that the solution can no longer be 
derived in such a straightforward manner. 

\section{The $D0-D6$ system}

The $D0-D6$ system is peculiarly different from other D-brane bound states in type $IIA$ theory. Here
we will first summarize the leading order result \cite{Nampuri:2007gv} and subsequently find the attractor solution in the 
presence of sub-leading corrections. 

\subsection{The leading order solution}

Unlike the $D0-D4$ bound state (along with $D0-D4-D6$ and $D0-D2-D4-D6$ bound states), the 
$D0-D6$ bound state does not admit supersymmetric solution in the absence of a $B$-field. 
One can see this in the following \cite{Witten:2000mf}{\footnote{The D-brane bound states in the 
presence of  $B$ field was discussed in \cite{Mihailescu:2000dn}}}:

For a supersymmetric $D0$ brane there must exist spinors $\epsilon^\alpha$ and $\hat\epsilon_\beta$
satisfying $\hat\epsilon_\beta = \Gamma^0_{\beta\alpha}\epsilon^\alpha$, where as a  $D6$ brane 
preserving supersymmetry need to satisfy the spinor condition 
$\hat\epsilon_\beta = (\Gamma^0\Gamma^1\cdots\Gamma^6)_{\beta\alpha}\epsilon^\alpha$. 
For the supersymmetric $D0-D6$ bound state to exist both the conditions must be satisfied 
simultaneously and hence the spinor $\epsilon$ will necessarily have to obey 
$\Gamma^1\cdots\Gamma^6\epsilon = \epsilon$. Since  $(\Gamma^1\cdots\Gamma^6)^2 = -1$,
this equation can never admit a nonzero  solution for the spinor $\epsilon$.

This can also be seen from the supergravity analysis by solving the condition $\nabla_aW=0$. We will
consider the known solution for supersymmetric  $D0-D4-D6$ black hole \cite{Maldacena:1997de} and 
show that the limit in which the $D4$ charges $p^a\rightarrow 0$ does not give a consistent solution. 
The supersymmetric $D0-D4-D6$ solution is given by:
\begin{eqnarray}
\label{susyd6}
x^a  =  p^a\frac {1}{ 2D}\left(p^0q_0 \pm  i \sqrt{q_0(4D -(p^0)^2 q_0)}\right)
\end{eqnarray}
We can see that, in the limit $p^a\rightarrow 0$ the imaginary part ${\Im}(x^a) = 0$ and hence the solution lies out side the K\"ahler cone.

Though the equation $\nabla_a W = 0$ can't be satisfied for $D0-D6$ system, there is no obstruction for 
solving $\partial_a V = 0$. We will now show that this equation has a physical solution by taking the 
limit $p^a\rightarrow 0$ in the non-supersymmetric $D0-D4-D6$ solution. Let us consider the explicit 
expression for the non-supersymmetric $D0-D4-D6$ solution \cite{Tripathy:2005qp}. In this case, the 
solution is naturally given in terms of a parameter $s = \sqrt{(p^0)^2-\frac{4D}{ q_0}}$. There exists two
branches in the charge lattice $|s/p^0| > 1$ as well as $|s/p^0|<1$ for which we have the solution, where 
as the line $|s/p^0|=1$ corresponds to the singular limit $p^a\rightarrow 0$.

To obtain the leading order results for  $D0-D6$ solution, we will consider the $D0-D4-D6$ solution 
in the region $|s/p^0|>1$ and take the limit $p^a\rightarrow 0$. It can be easily verified that the same 
result can be obtained  by considering the limit $p^a \rightarrow 0$ in the region $|s/p^0|<1$. The 
non-supersymmetric solution for $D0-D4-D6$ system valid in the region $|s/p^0|>1$ is given by 
\cite{Tripathy:2005qp}
$$x^a = p^a (t_1 + i t_2) \ , $$ 
where $t_1$ and $t_2$ have the following form 
\begin{eqnarray}
\label{t1d6}
t_1 & =  & 2\  \frac{(s+p^0)^{1/3} - (s-p^0)^{1/3}}{(s+p^0)^{4/3} + (s-p^0)^{4/3}} \\
t_2 &= &  \frac{4 s}{ (s^2-(p^0)^2)^{1/3} \left((s+p^0)^{4/3}+(s-p^0)^{4/3}\right)}
\end{eqnarray}
We see that $t_1$ remains finite and hence the real part of $x^a$ vanishes  in the limit 
$p^a\rightarrow  0$. On the other hand $t_2$ diverges in this limit and the resulting solution lies inside 
the K\"ahler cone. We find:
\begin{eqnarray}
\label{d0d6}
\lim_{p^a \to 0} \ x^a = \lim_{p^a\to 0} \ i \ \frac{ p^a }{2}\left(-\frac{q_0}{D p^0}\right)^{1/3}
= i \hat x_\mp^a \left(\pm\frac{q_0}{p^0}\right)^{1/3} \ ,
\end{eqnarray}
where  \[\hat x_\pm^a = \lim_{p^a\to 0} (\pm p^a/D^{1/3})\ . \] 
Here $\hat x^a_+$ and $\hat x^a_-$ are two real vector restricted to the hypersurfaces 
$D_{abc}\hat x^a_+\hat x^b_+\hat x^c_+ = 1$ and $D_{abc}\hat x^a_-\hat x^b_-\hat x^c_- = -1$
 in the moduli space respectively. There are two branches of solutions for the $D0-D6$ attractor:
\begin{eqnarray}\label{d0d6n}
x^a = \left\{\begin{matrix} & i\hat x_-^a \left(\frac{q_0}{p^0}\right)^{(1/3)} \ {\rm For}\  q_0p^0 > 0 \ ,  \cr 
& i\hat x_+^a \left(-\frac{q_0}{p^0}\right)^{(1/3)}  \ {\rm For}\  q_0p^0<0 \ . \   \end{matrix} \right.
\end{eqnarray}
The entropy of the black hole, for both the cases, is given by $S = \pi |q_0p^0|$.
 Note that the  solutions (\ref{d0d6n}) obey
\begin{equation}
D_{abc} x^a x^b x^c = i \frac{q_0}{8 p^0} \ .
\end{equation}
This equation involves $n$ real variables and hence defines a $(n-1)$ dimensional hypersurface 
in the $2n$ dimensional moduli space. Thus we find that the moduli are not completely fixed on the 
black hole horizon and  the space of attractor points is specified by a $(n-1)$ dimensional 
hypersurface in the moduli space.

\subsection{The sub-leading correction}

We will now consider the $D0-D6$ system in the presence of perturbative sub-leading corrections.
The superpotential (\ref{suppot}) for the system takes the form 
\begin{equation}
W = q_0 - p^0 \alpha_{0a} x^a + p^0 D_{abc} x^a x^b x^c \ , 
\end{equation}
where as the K\"ahler potential is given by eq.(\ref{IIAkp}). Unlike the $D0-D4$ system, here we 
can't make a simple redefinition of charges to absorb the effect of sub-leading corrections and
need to explicitly solve the equations of motion. We will substitute these expression for $W$ and 
$K$ in 
eq.(\ref{critpt})
\begin{equation}
 \left(g^{b\bar c} \nabla_a\nabla_b W \overline{\nabla_cW} + 2 \nabla_a W \overline{W}
+ \partial_a g^{b\bar c} \nabla_b W\overline{\nabla_c W}\right) = 0~,
\label{newcritpt}
\end{equation}
and solve it explicitly. Taking a clue from the previous subsection we consider the following ansatz
for the scalar fields:
\begin{equation}
x^a = {\hat{x}}^a t \ ,
\label{newansatz}
\end{equation} 
for some real vector $\hat x^a  $. 

Before we proceed to solve eq.(\ref{newcritpt}), we will try to solve the  supersymmetry condition
 $\nabla_a W = 0$ and show that this equation does not admit any physically acceptable solution in the 
 large volume limit. The leading order result was discussed in the previous sub-section and the world 
 sheet analysis for nonexistence of the $D0-D6$ bound state  was presented in Ref.\cite{Witten:2000mf}.
After a straightforward computation, we find 
\begin{equation}
 \nabla_aW=\hat{D}_a\left(3t^2+\frac{3iW}{2 p^0 \hat{D}t_2}\right)-\alpha_{0a} = 0 \ ,
  \end{equation}
  Here for convenience we introduce $\hat D_{ab} = D_{abc}\hat x^c, \hat{ D}_a=\hat{D}_{ab}\hat{x}^b$ 
  and $\hat{D} =\hat{ D}_a \hat{x}^a$. The inverse of the matrix $\hat D_{ab}$ is denoted by $\hat D^{ab}$.
  We also define, $\hat\alpha_0 = \alpha_{0a}\hat x^a$ and set $t = t_1 + i t_2$. 
  Now, equating the real as well as the imaginary parts of the above equation to zero separately, we find
  \begin{eqnarray}
  3 \hat{D}({t_1}^2+{t_2}^2)- \hat{\alpha}_0 &=&0 \cr
  q_0+p^0\hat{D}t_1({t_1}^2+{t_2}^2)-p^0t_1\hat{\alpha}_0&=&0 
  \label{leadingsusy}
  \end{eqnarray}
  Solving the above equations for $t$ we get
 \begin{equation}
  t=\frac{3q_0}{2p^0\hat{\alpha}_0} + \frac{i}{2}\sqrt{\frac{4\hat{\alpha}_0}{3\hat{D}}- \frac{9q_0^2}{({\hat{\alpha}_0} p^0)^2}} \ .
  \end{equation}
  
We will focus on the imaginary part of $t$, since this is the quantity which appears in the expression for the 
K\"ahler potential and hence in the volume of the Calabi-Yau manifold. Apart from the charges $q_0$ and 
$p^0$ it depends on $\hat{\alpha}_0$ and $\hat{D}$, which involve  the intersection numbers, weight of projective 
coordinates, degree of hypersurfaces, etc, and are $O(1)$ quantities. Since $\hat x^a$ is invariant under the 
scaling $\hat x^a\to \lambda\hat x^a$, we can't make its imaginary part arbitrarily large by choosing $\hat x^a$  
of arbitrarily small size. Hence the imaginary part for this solution, if at all it exists for some Calabi-Yau 
manifold, can at best be of order $O(1)$. Thus the solution  (\ref{leadingsusy}) is not valid in the large 
volume limit we are considering.

We will now turn into the exact solution for the non-supersymmetric attractor. Each of the terms in 
eq.(\ref{newcritpt}) will be evaluated for the above ansatz in the appendix. Adding them together we find,
after some simplification
\begin{eqnarray}
\label{dav}
0 & = & \frac{3i}{\hat{D}t_2}\left(2{q_0}^2+2(p^0)^2\hat{D}^2\bar{t}^4t^2+4p^0q_0 t_1(\hat{D}t_1\bar{t}-\hat{\alpha}_0)\right.\nonumber\\&-&\left.4(p^0)^2\hat{D}t_1\bar{t}^2t\hat{\alpha}_0+(2{t_1}^2+{t_2}^2)(p^0)^2{\hat{\alpha}_0}^2)\right)\hat{D}_a\cr &+&\left(4p^0t_1\hat{\alpha}_0-4q_0
-2ip^0t_2\hat{\alpha}_0-4\hat{D}p^0t_1\bar{t}^2\right)p^0\hat{\alpha}_{0a}-\frac{i\hat{D}(p^0)^2t_2}{3}\hat T_a \nonumber\ ,
\end{eqnarray}
where $\hat{T}_a=D_{apq}\hat{D}^{pb}\hat{D}^{qc}\alpha_{0b}\alpha_{0c}$.
The real part of the above equation gives
\begin{eqnarray}
0 =3t_1(q_0t_1+p^0|t|^2(\hat{D}|t|^2-\hat{\alpha}_0))\hat{D}_a
-\left(q_0+p^0t_1(\hat{D}{t_1}^2-\hat{D}{t_2}^2-\hat{\alpha}_0)\right)\alpha_{0a}
\end{eqnarray}
Defining $L$ to be 
\begin{equation}
\label{M}
 L=\frac{3t_1(q_0t_1+p^0|t|^2(\hat{D}|t|^2-\hat{\alpha}_0))}{q_0+p^0t_1(\hat{D}{t_1}^2-\hat{D}{t_2}^2-\hat{\alpha}_0)} \ ,
 \end{equation}
 we find 
 \begin{equation}
\label{Da}
\alpha_{0a}=\hat{D}_aL
\end{equation}
The vectors $\hat{D}_a$ and $\alpha_{0a}$ depend on the property of the Calabi-Yau manifod. In addition,
$\hat{D}_a$ depends on $\hat{x}^a$ as well. A priory there is no reason why both these vectors will be aligned 
in the same direction and hence this equation imposes restriction on $\hat{x}^a$. Multiplying $\hat{x}^a$ on
both sides and summing over $a$ we find $L = \hat{\alpha}_0/\hat{D}$ and $\hat{D}\alpha_{0a} = \hat{D}_a\hat{\alpha}_0$. The later of imposes restriction on $\hat x^a$ and hence defines the hypersurface of 
attractor points in presence of sub-leading terms. 

The relation $L = \hat\alpha_0/\hat D$ gives
\begin{equation}
\label{M1}
 \frac{\hat{\alpha}_0}{\hat{D}}=\frac{3t_1(q_0t_1+p^0|t|^2(\hat{D}|t|^2-\hat{\alpha}_0))}{q_0+p^0t_1(\hat{D}{t_1}^2-\hat{D}{t_2}^2-\hat{\alpha}_0)} \ ,
 \end{equation}
This equation, along with the imaginary part of eq.(\ref{dav}) can be simultaneously solved to get $t$ in 
terms of the charges $q_0, p^0$ and the geometric quantities $\hat{}\alpha_0, \hat{D}$.
The algebraic equations, we get from eq.(\ref{dav}) in this process, are given as 
\begin{eqnarray}
\label{E1}
0&=&36{q_0}^2+36(p^0\hat{D})^2({t_1}^2+{t_2}^2)^2({t_1}^2-{t_2}^2)+6(p^0)^2(6{t_1}^2+{t_2}^2){\hat{\alpha}_0}^2\nonumber\\&+&72q_0p^0t_1(\hat{D}{t_1}^2-\hat{\alpha}_0)-2(p^0)^2\hat{D}(\hat{T}{t_2}^2+12{t_1}^2(3{t_1}^2+{t_2}^2)\hat{\alpha}_0) \ , 
\end{eqnarray}and
\begin{eqnarray}
\label{E2}
3p^0\hat{D}^2t_1({t_1}^2+{t_2}^2)^2+\hat{\alpha}_0(p^0t_1\hat{\alpha}_0-q_0)+\hat{D}t_1(3q_0t_1-2(2{t_1}^2+{t_2}^2)p^0\hat{\alpha}_0)=0 \ \  
\end{eqnarray}
We can rescale the variables to write these two equations in a simple form. The resulting equations depend
 on one parameter only. The details are worked out in \S{A.1}.  Subsequently, we can eliminate $t_1$ and 
 $t_2$  in sequence to get two cubic equations, which we can solve exactly. The resulting solution is  
\begin{eqnarray}
t_1 &=& \frac{q_0}{\hat\alpha_0 p^0} \hat F_1\left(\frac{q_0^2\hat D}{\hat\alpha_0^3(p^0)^2}\right)\ , \\
t_2^2 &=& \frac{\hat\alpha_0}{\hat D} \hat F_2\left(4 - 27 \frac{\hat D q_0^2}{\hat\alpha_0^3(p^0)^2}\right) \ ,
\end{eqnarray}
where the functions $\hat F_1(x)$ and $\hat F_2(x)$ are defined to be 
\begin{eqnarray}
\hat F_1(x) &=& \frac{1}{6 x (27 x - 2)} \left( 18 x + \hat f_+(x)^{(1/3)} - \hat f_-(x)^{(1/3)}\right) \\
\hat F_2(x) &=& \frac{1}{6 (x - 2)^2}\left(\hat g_+(x)^{(1/3)} - \hat g_-(x)^{(1/3)} - 6 x\right) \ .
\end{eqnarray}
The functions $\hat f_\pm(x)$  and $\hat g_\pm(x)$ in turn are given by
\begin{eqnarray}
\hat f_{\pm}(x) &=& 12\left((2 - 27 x) \sqrt{3 x^3 (27 x - 4)^3} \pm 9 x^2 (27 x - 4)^2\right) \\
\hat g_\pm(x) &=& 4 \left((x-2)^3 \sqrt{x^7 (x-4)} \pm (2 x^3 - 40 x^4 - 6 x^5 + 8 x^6 - x^7)\right) \ .
\end{eqnarray}
Substituting the solution for $t = t_{1} + i t_{2}$ in  the entropy $S=\pi V(\hat x_0 t)$, we find, for  the 
non-supersymmetric attractor
\begin{equation}
S=\frac{\pi p^{0}\hat{\alpha}_{0}}{3}\sqrt{\frac{9{q_{0}}^2}{(p^{0})^{2}{\hat{\alpha}_{0}}^{2}}-\frac{4\hat{\alpha}_{0}}{3\hat{D}}}
\end{equation}

   \subsection{The zero modes}

We saw that the leading $D0-D6$ solution has $(n-1)$ zero modes. We are interested in studying the 
effect of sub-leading corrections to these zero modes. We will address this issue in this section. We 
need to consider the  fluctuations around $\hat x^a t$:
$$x^a  = \hat x^a t + \delta \xi^{a} + i \delta y^{a}$$
and keep term quadratic in the fields $\delta\xi^a, \delta y^a$ in the effective black hole potential:
\begin{eqnarray}
V_{quad} & = &  \partial_a\partial_{\bar d} V (\delta \xi^{a}\delta \xi^{d} + \delta y^{a} \delta y^{d}) +  \Re(\partial_a\partial_d V)
(\delta \xi^{a}\delta \xi^{d} - \delta y^{a} \delta y^{d}) \cr && - 2\Im(\partial_a\partial_d V)\delta \xi ^{a} \delta y^{d}.
\end{eqnarray} 
Here $\Re(\partial_a\partial_d V)$ and $\Im(\partial_a\partial_d V)$ are the real and imaginary 
parts of $(\partial_a\partial_d V)$ respectively. Each of these terms are computed explicitly in \S{A.2}. 
Following \cite{Tripathy:2005qp}, we express the mass matrix $M$ in the form:
\begin{eqnarray}
  \label{massmatrixapp}
M=E \left(3\frac{\hat{D}_a \hat{D}_d}{  \hat{D}}-\hat{D}_{ad}\right) \otimes {\bf I} +  \hat{D}_{ab} \otimes (A \sigma^3 - B  \sigma^1),
  \end{eqnarray}  
The coefficients $E, A$ and $B$ appearing in the above formula are computed in \S{A.2} and are given by 
\begin{eqnarray}
E &=&e^{K_0}\left(\frac{4{\hat{\alpha}_0}^2(p^0)^2(3-u)(u+3)^2}{3\hat{D}(u-1)^2}\right) \cr
A &=& e^{K_0}\left(\frac{2{\hat{\alpha}_0}^2(p^0)^2(3-u)(u+3)^2}{3\hat{D}(u-1)^2}(u^2-2u-1)\right) \cr
B &=&  e^{K_0}\left(\frac{2{\hat{\alpha}_0}^2(p^0)^2(3-u)(u+3)^2}{3\hat{D}(u-1)}\sqrt{(u+1)(3-u)})\right) 
\end{eqnarray}
Here $K_0$ is the K\"ahler potential evaluated at the attractor point and the parameter $u$ is given by 
$u = \hat F_3(\hat D q_0^2/(\hat\alpha_0^3(p^0)^2))$. The function $\hat F_3(x)$ is defined as
\begin{equation}
\hat F_3(x) = \sqrt{1 - 24 x (\hat F_1(x))^2} \ . 
\end{equation}
Substituting the explicit expression for $A$ and $B$, we can see that the $2\times 2$ matrix 
$(A\sigma^3 - B  \sigma^1)$ has the Eigenvalues $\pm E$. Thus the matrix $M$ can be brought 
into block diagonal form, where one block contains a positive coefficient times the moduli space 
metric $g_{a\bar b}$ at the extremum and the other block contains a positive coefficient times 
the matrix $D_a D_b$. Since the moduli space metric is positive definite and any matrix of the 
form $D_aD_b$ has one positive and $(n-1)$ zero eigenvalues,  the matrix $M$ has  $(n+1)$ 
positive  and $(n-1)$ zero eigenvalues. 

\section{Conclusion}

In this paper we have studied non-supersymmetric attractors in type $IIA$ string theory compactified on
a Calabi-Yau manifold, in  the presence of perturbative sub-leading corrections to the perpotential. We 
discussed both $D0-D4$ as well as $D0-D6$ configurations. In both the cases there are $(n-1)$ massless 
modes. Because of the presence of non-trivial quartic terms, in  the case of $D0-D4$ black holes, all 
the vector multiplet moduli are attracted to a fixed point. On the other hand, for the $D0-D6$ system there
exists a $(n-1)$ dimensional hypersurface of attractor points.  Thus, in this case the $(n-1)$ massless modes are 
exactly flat and the attractor mechanism occurs only in an $(n+1)$ dimensional sub-space of the $2 n$ 
dimensional moduli space for the vector multiples moduli. The quantum correction deforms this $(n-1)$
dimensional  hypersurface but does not change its dimensionality. Since the perturbative corrections are not 
sufficient to lift the massless modes, they can only be lifted by non-perturbative terms in the potential. 
It would be interesting to consider explicit examples where these modes are stabilized by non-perturbative
corrections. It would also be interesting to find the full flow and understanding it in terms the first 
order formalism. We hope to report on some of these issues in future.

\section{Acknowledgments}
We would like to thank S. Govindarajan and I. Karthik for helpful discussions.  This work was partially supported by the Indo-French Centre for the Promotion of Advanced Research (CEFIPRA) Project No. 4104-2.

\section{Note Added}
The effect of sub-leading corrections in $st^2$ and $t^3$ models has been studies in 
\cite{Bellucci:2007eh,Bellucci:2008tx}. The effect of perturbative 
quantum corrections on mass less moduli and on flat directions had already been been 
obtained in \cite{Bellucci:2010zd} using symplectic transformations. 
We are grateful to Alessio Marrani for pointing out the above references to us.

\appendix

\section{Appendix}
 
 In this appendix we will carry out some of the computations in detail. In \S{A.1} we will derive the attractor
 solution for $D0-D6$ system and in \S{A.2} we will outline some of  the steps involved in obtaining mass
 matrix for this system.
 
\subsection{Nonsupersymmetric  solution for $D0-D6$ system}
The black hole effective potential for $N=2$ supergravity in four dimensions is given by 
\begin{eqnarray}
\label{aveff}
V&=&e^K\left[g^{a\bar{b}}\nabla_aW\left(\nabla_bW\right)^*+|W|^2\right]
\end{eqnarray}
where $\nabla_aW=\partial_a W + \partial_aK W$ . 
In terms of the $N=2$ prepotential $F$, the K\"ahler potential $K$ and superpotential $W$ are given by
\begin{eqnarray}
K &=& - \ln{\Im}\left(\sum_{a=0}^N\bar{X}^{\bar a}\partial_a F(X)\right) \cr
W &=& \sum_{a=0}^N(q_a X^a - p^a \partial_a F) \ .
\label{akahle}
\end{eqnarray}
We are interested in finding the non-supersymmetric attractors, which  correspond to extremising this potential.
The stable non-supersymmetric attractors correspond to the minima of $V$ for which $\nabla_aW\neq 0$. The
equation of motion is 
\begin{equation}\label{aveffprim}
 \left(g^{b\bar c} \nabla_a\nabla_b W \overline{\nabla_cW} + 2 \nabla_a W \overline{W}
+ \partial_a g^{b\bar c} \nabla_b W\overline{\nabla_c W}\right) = 0~.
\end{equation}

This section deals with the attractor solution for the $D0-D6$ system. We will first evaluate each term in the 
equation of motion separately. Subsequently we will add them up and simplify and eventually find the exact
solution corresponding to the equation of motion.

We will first introduce some of the standard notations and express the K\"ahler potential, moduli space metric 
and its inverse in terms of them \cite{Tripathy:2005qp}:
\begin{eqnarray}
\label{aeqm}
&& M_{ab} =  D_{abc} (x^c - \bar x^c) \cr
&& M_a =  D_{abc} (x^b - \bar x^b) (x^c - \bar x^c) \cr
&& M = D_{abc} (x^a - \bar x^a) (x^b - \bar x^b) (x^c - \bar x^c) ~.
\end{eqnarray}
The K\"ahler potential is 
\begin{equation}
K = - \ln(-i M)
\end{equation}
The metric $g_{a\bar b}=\partial_a\partial_{\bar b} K$ and its inverse are 
\begin{eqnarray}
g_{a\bar b} &= &\frac{3}{M} \left( 2 M_{ab} - \frac{3}{ M} M_a M_b\right)~, \cr
g^{a\bar b} &=& \frac{M}{6} \left( M^{ab} -\frac {3}{ M} (x^a - \bar x^a)
(x^b - \bar x^b) \right) ~,
\end{eqnarray}
We will also need 
\begin{eqnarray}
\partial_ag^{b\bar{c}}=\frac{1}{6}\left(3M_aM^{bc}-MM^{pb}M^{qc}D_{apq}\right)-\frac{1}{2}\left({\delta_a}^b(x^c - \bar x^c)+{\delta_a}^c(x^b - \bar x^b)\right)
\end{eqnarray}

For $D0-D6$ system, the superpotential $W$ and its covariant derivatives $\nabla_aW, \nabla_a\nabla_bW$
are given by
\begin{eqnarray}
W&=&q_0-p^0\alpha_{0a}x^a+p^0D_{abc}x^a x^b x^c \cr
\nabla_aW&=&-p^0\alpha_{0a}+3p^0D_{abc}x^b x^c-\frac{3M_aW}{M}\nonumber\\
\nabla_a\nabla_bW&=&6p^0D_{abc}x^c+\frac{6W}{M}\left(\frac{3M_aM_b}{M}-M_{ab}\right) \cr
&+&\frac{3p^0}{M}\left(M_a\alpha_{0b}+M_b\alpha_{0a}\right) 
-\frac{9p^0x^px^q}{M}\left(M_aD_{bpq}+M_bD_{apq}\right)
\end{eqnarray}
We will use the ansatz $x^a=\hat{x}^at=\hat{x}^a(t_1+it_2)$. The inverse metric and its derivative in this ansatz
has the form
\begin{eqnarray}
g^{b\bar c} &=&\frac {2 t_2^2}{ 3}\hat{ D} \left(\frac {3}{\hat{D}} \hat{x}^b\hat{ x}^c - \hat{D}^{bc}\right) \ , \cr
\partial_a g^{b\bar c} &=& -\frac{i t_2}{ 3} \hat{D} \left( \frac{3}{ \hat{D}} (\hat{x}^c \delta_a^b
+ \hat{x}^b\delta^c_a - \hat{D}^{bc} \hat{D}_a) + \hat{D}^{ec} \hat{D}^{bf} D_{aef}\right) \ ,
\end{eqnarray}
where as the superpotential $W$ and its covariant derivatives are given by
\begin{eqnarray}
W&=&q_0+p^0\hat{D}t^3-p^0\hat{\alpha}_0 t\nonumber\\
 \nabla_aW&=&\hat{D}_a p^0 \left(3t^2+\frac{3iW}{2p^0\hat{D}t_2}\right)-p^0\hat{\alpha}_{0a} \cr
 \nabla_a\nabla_bW&=&\hat{D}_{ab}\left(6p^0t+\frac{3W}{2\hat{D}{t_2}^2}\right) \cr
 &+&\hat{D}_a\hat{D}_b\left(\frac{-9W}{2\hat{D}^2{t_2}^2}+\frac{9ip^0t^2}{\hat{D}t_2}\right)
+\frac{3p^0}{2i\hat{D}t_2}\left(\hat{D}_a\alpha_{0b}+\hat{D}_b\alpha_{0a}\right)
\end{eqnarray}

Let us  evaluate each  term in eq.(\ref{aveffprim}).
\begin{eqnarray}
\label{term1}
 g^{b\bar c} \nabla_a\nabla_b W \overline{\nabla_cW}&=&\hat{D}_a\left(m_1\left(6p^0t-\frac{3W}{\hat{D}{t_2}^2}+\frac{9ip^0t^2}{t_2}+\frac{3p^0\hat{\alpha}_0}{2i\hat{D}t_2}\right)\right.\nonumber\\
 &+&\left.m_2\left(\frac{-9W \alpha}{2\hat{D}^2{t_2}^2}+\frac{9ip^0t^2\hat{\alpha}_0}{\hat{D}t_2}+\frac{3p^0\hat{T}}{2i\hat{D}t_2}\right)\right)\nonumber\\
 &+&\alpha_{0a}\left(\frac{3p^0m_1}{2it_2}+m_2\left(6p^0t+\frac{3W}{2\hat{D}{t_2}^2}+\frac{3p^0\alpha_0}{2i\hat{D}t_2}\right)\right)
  \end{eqnarray}
Where $m_1=\frac{2p^0\hat{D}{t_2}^2}{3}\left(2\bar{G}-\frac{3\hat{\alpha}_0}{\hat{D}}\right)$, ~~$m_2=\frac{2p^0\hat{D}{t_2}^2}{3}$,~~ $G=3t^2+\frac{3iW}{2p^0\hat{D}t_2}$ \\
and $\hat{T}=\hat{D}^{bc}\alpha_{0b}\alpha_{0c}$
\begin{eqnarray}
\label{term2}
\partial_a g^{b\bar c} \nabla_b W\overline{\nabla_c W}&=&\hat{D}_a\left(it_2\hat{T}-\frac{4}{3}it_2\hat{D}G\bar{G}\right)(p^0)^2-\frac{it_2\hat{D}(p^0)^2}{3}\hat{T}_a\nonumber\\
&+&\alpha_{0a}\left(\frac{4}{3}it_2\hat{D}(G+\bar{G})-2it_2\hat{\alpha}_0\right)(p^0)^2
\end{eqnarray}
Where $\hat{T}_a=D_{apq}\hat{D}^{pb}\hat{D}^{qc}\alpha_{0b}\alpha_{0c}$
\begin{eqnarray}
\label{term3}
2 \nabla_a W \overline{W}&=&2p^0\left(\hat{D}_aG-\alpha_{0a}\right)\overline{W}
\end{eqnarray}
Adding up terms eq. (\ref{term1}), eq. (\ref{term2}), eq. (\ref{term3}) and using eq.(\ref{aveffprim}) we get,
\begin{eqnarray}
\label{adav}
0 & = & \frac{3i}{\hat{D}t_2}\left(2{q_0}^2+2(p^0)^2\hat{D}^2\bar{t}^4t^2+4p^0q_0 t_1(\hat{D}t_1\bar{t}-\hat{\alpha}_0)\right.\nonumber\\&-&\left.4(p^0)^2\hat{D}t_1\bar{t}^2t\hat{\alpha}_0+(2{t_1}^2+{t_2}^2)(p^0)^2{\hat{\alpha}_0}^2)\right)\hat{D}_a\cr &+&\left(4p^0t_1\hat{\alpha}_0-4q_0
-2ip^0t_2\hat{\alpha}_0-4\hat{D}p^0t_1\bar{t}^2\right)p^0\hat{\alpha}_{0a}-\frac{i\hat{D}(p^0)^2t_2}{3}T_a \nonumber\ ,
\end{eqnarray}
         where $\hat{T}_a=D_{apq}\hat{D}^{pb}\hat{D}^{qc}\alpha_{0b}\alpha_{0c}$.
Taking the real part of the above equation we get.
\begin{eqnarray}
\Re(e^K\partial_aV_{eff})&=&\left(3t_1(q_0t_1+p^0|t|^2(\hat{D}|t|^2-\hat{\alpha}_0)\right)p^0\hat{D}_a\nonumber\\&+&\left(-q_0+p^0t_1(-\hat{D}{t_1}^2+\hat{D}{t_2}^2+\hat{\alpha}_0)\right)p^0\alpha_{0a}=0
\end{eqnarray}
This gives 
\begin{equation}
\label{aDa}
\alpha_{0a}=\hat{D}_aL
\end{equation}
Where
\begin{equation}
\label{aM}
 L=\frac{3t_1(q_0t_1+p^0|t|^2(\hat{D}|t|^2-\hat{\alpha}_0))}{q_0+p^0t_1(\hat{D}{t_1}^2-\hat{D}{t_2}^2-\hat{\alpha}_0)}.
 \end{equation}
 
 Multiplying  eq.(\ref{adav}) with $\hat x^a$ and separate its real and imaginary parts we get the following two equations.
\begin{eqnarray}
\label{aE1}
0&=&36{q_0}^2+36(p^0\hat{D})^2({t_1}^2+{t_2}^2)^2({t_1}^2-{t_2}^2)+6(p^0)^2(6{t_1}^2+{t_2}^2){\hat{\alpha}_0}^2\nonumber\\&+&72q_0p^0t_1(\hat{D}{t_1}^2-\hat{\alpha}_0)-2(p^0)^2\hat{D}(\hat{T}{t_2}^2+12{t_1}^2(3{t_1}^2+{t_2}^2)\hat{\alpha}_0) \ , 
\end{eqnarray}and
\begin{eqnarray}
\label{aE2}
3p^0\hat{D}^2t_1({t_1}^2+{t_2}^2)^2+\hat{\alpha}_0(p^0t_1\hat{\alpha}_0-q_0)+\hat{D}t_1(3q_0t_1-2(2{t_1}^2+{t_2}^2)p^0\hat{\alpha}_0)=0 \ \ \  
\end{eqnarray}
We need to solve these two equations for $t_1$ and $t_2$ in terms of $q_0, p^0,\hat D$ and $\hat\alpha_0$.
We will first do some scaling of variables and parameters to simplify these two equations. Introducing the 
variables $\tilde t_1 = t_1 (p^0\hat\alpha_0 /q_0)$ and $\tilde t_2 = t_2 \sqrt{\hat D/\hat\alpha_0} $, we find
\begin{eqnarray}
9 \hat D^2 (p^0)^2 q_0^4\hat\alpha_0^3  {\tilde t_1}^3 (2 + \tilde t_1 ({\tilde t_2}^2 - 2)) 
+ (p^0)^6 \hat\alpha^9 {\tilde t_2}^2 ( 1 - 9 {\tilde t_2}^4) && \cr + 9 \hat D^3 q_0^6 {\tilde t_1}^6 
+ 3 \hat D (p^0)^4 q_0^2 \hat\alpha_0^6 (3 - 6{\tilde t_1} + {\tilde t_1}^2 (3 - 2{\tilde t_2}^2 - 3 {\tilde t_2}^4))
& = & 0
\end{eqnarray}
and
\begin{eqnarray}
3 \hat D^2 q_0^4 {\tilde t_1}^5 
+ \hat D (p^0)^2 q_0^2 \hat\alpha_0^3  {\tilde t_1}^2 (3 + \tilde t_1 (6\tilde t_2^2 - 4))-(p^0)^4 \hat\alpha_0^6 (1 - \tilde t_1(1- 2  \tilde t_2^2 + 3\tilde t_2^4)) = 0 \ \ \ 
\end{eqnarray}
We will now introduce the parameter $\tilde D =  \hat D q_0^2/((p^0)^2\hat\alpha_0^3)$. The above equations 
take particularly simple form when expressed in terms of $\tilde D$:
\begin{eqnarray}
9 \tilde D (1 - \tilde t_1 + \tilde D \tilde t_1^3)^2 + (1 - 3 \tilde D \tilde t_1^2)^2 \tilde t_2^2 - 9 \tilde D \tilde t_1^2 \tilde t_2^4 - p \tilde t_2^6=0 \\
1 - \tilde t_1 (1 + 3\tilde D^2 \tilde t_1^4 - 2 \tilde t_2^2 + 3 \tilde t_2^4 + \tilde D \tilde t_1 (3 + \tilde t_1 (6 \tilde t_2^2 - 4))) = 0
\end{eqnarray}
Eliminating $\tilde t_1$ and $\tilde t_2$ in succession, and after a bit simplification, we find
\begin{eqnarray}
\tilde D (27 \tilde D - 2) \tilde t_1^3  - 9 \tilde D \tilde t_1^2 + 2 \tilde t_1 - 1 = 0 \\
27 (\tilde d - 2)^2 \tilde t_2^6 + 81 \tilde d \tilde t_2^4 + 18 \tilde d^2 \tilde t_2^2 + \tilde d^3 = 0
\end{eqnarray}
Here, for easy reading, we have introduced $\tilde d = 4 - 27 \tilde D$ in the second line. We finally get two
cubic equations in terms of variables $\tilde t_1$ and $\tilde t_2^2$ which we can solve easily. The exact 
solution for the original variables $t_1,t_2$ in terms of the charges $q_0,p^0$ and parameters $\hat D,
\hat\alpha_0$ is given in \S4.2.\\
We will now find the black hole entropy. The Entropy of the non-supersymmetric solution determined by the 
value of the black hole  effective potential at the critical point, $S=\pi V(\hat x_0 t)$.
 In terms of $t_{1}$ and $t_{2}$ we can write the effective potential has the following form:
 \begin{eqnarray}
 V=\frac{3{q_{0}}^{2}+3\hat{D}^{2}|t|^{6}-6\hat{D}{t_{1}}^{2}|t|^{2}\hat{\alpha}_{0}+(3{t_{1}}^{2}+{t_{2}}^{2}){\hat{\alpha}_{0}}^{2}+6q_{0}(\hat{D}{t_{1}}^{3}-t_{1}\hat{\alpha}_{0})}{6\hat{D}{t_{2}}^{3}}
 \end{eqnarray}
 Substituting the solution for $t_1$ and $t_2$ in the above expression,  we get the entropy of the non-supersymmetric solution:
 \begin{equation}
S=\frac{\pi p^{0}\hat{\alpha}_{0}}{3}\sqrt{\frac{9{q_{0}}^2}{(p^{0})^{2}{\hat{\alpha}_{0}}^{2}}-\frac{4\hat{\alpha}_{0}}{3\hat{D}}}
\end{equation}

\subsection{The mass matrix}

In this section we will  compute the mass matrix for the $D0-D6$ system. We need the coefficients of the quadratic 
terms in the effective potential.  It is straightforward to express them in terms of $W$ and its 
covariant derivatives \cite{Tripathy:2005qp}:
\begin{eqnarray}
\label{atwoderiv}
e^{-K_0} \partial_a\partial_d V &=& \left\{ 
g^{b\bar c} \nabla_a\nabla_b\nabla_d W 
+ \partial_a g^{b\bar c} \nabla_b\nabla_d W + \partial_d g^{b\bar c} \nabla_b\nabla_a W
\right\} \overline{\nabla_cW} \cr &+&
 3 \nabla_a\nabla_d W \overline{W}
+ \partial_a\partial_d g^{b\bar c} 
\nabla_b W \overline{\nabla_cW}
- g^{b\bar c}\partial_a g_{d\bar c} \nabla_b W \overline{W} \cr
e^{-K_0} \partial_a\partial_{\bar d} V &=&
g^{b\bar c} \nabla_a\nabla_b W\overline{\nabla_c \nabla_d W}
+ \left\{ 2 |W|^2 + g^{b\bar c}  \nabla_b W\overline{\nabla_c W}\right\} g_{a\bar d}
\cr &+&
 \partial_a g^{b\bar c} \nabla_b W \overline{\nabla_c \nabla_d W}
+ \partial_{\bar d} g^{b\bar c} \nabla_a\nabla_b W \overline{\nabla_c W}
+ 3 \nabla_a W\overline{\nabla_d W}
\cr &+&
  \partial_a\partial_{\bar d} g^{b\bar c} \nabla_b W \overline{\nabla_c W}
\end{eqnarray}
We need to explicitly evaluate these terms at the attractor point. Since the expressions are particularly 
lengthy, we will introduce some further notations and express various terms in eq.(\ref{atwoderiv}) in
terms of them in a compact way. Define:

\begin{eqnarray*}
g_1&=&6t+\frac{3W}{2p^0\hat{D}{t_2}^2}, 
\ \ g_2=\frac{-9W}{2p^0\hat{D}{t_2}^2}+\frac{9it^2}{t_2}+\frac{3L}{it_2}, \\
g_3&=&\frac{2{t_2}^2}{3}\left(3g_1+2g_2\right) , \ \ 
g_4=\frac{-2{t_2}^2g_1}{3}, \ \  g_5 = G-L,\ \ 
g_6 = 3t^2-L .
\end{eqnarray*}
The covariant derivatives of the superpotential, in terms of these quantities are 
\begin{eqnarray}
\nabla_aW&=&g_5\hat{D}_ap^0\nonumber\\
\nabla_a\nabla_bW&=&p^0\left(g_1\hat{D}_{ab}+g_2\frac{\hat{D}_a\hat{D}_b}{\hat{D}}\right)\nonumber\\
\nabla_a\nabla_bWg^{b\bar{c}}&=&p^0\left(g_3\hat{D}_a\hat{x}^c+g_4\hat{D}{\delta_a}^c\right)
\end{eqnarray}
We are now in a position to compute the mass matrix. Let us first consider individual  terms in $\partial_a\partial_{\bar d} V$ and simplify them. We find
\begin{eqnarray}
g^{b\bar c} \nabla_a\nabla_b W \overline{\nabla_c\nabla_dW} &=&(p^0)^2\left( \hat{D}_a\hat{D}_d\left(\bar{g_1}g_3+\bar{g_2}g_3+\bar{g_2}g_4\right)+\bar{g_1}g_4\hat{D}\hat{D}_{ad}\right)\nonumber\\
3 \nabla_aW\overline{\nabla_dW} &=& 3(p^0)^2g_5\bar{g_5}\hat{D}_a\hat{D}_d\nonumber\\
2 g_{a\bar d} |W|^2 &=& \frac{3}{\hat{D}{t_2}^2}\left(\frac{3\hat{D}_a\hat{D}_d}{2\hat{D}}-\hat{D}_{ad}\right)W\overline{W}\nonumber\\
g_{a\bar d} g^{b\bar c} \nabla_bW \overline{\nabla_cW} &=& (p^0)^2g_5\bar{g_5}\left(3\hat{D}_a\hat{D}_d-2\hat{D}\hat{D}_{ad}\right)\nonumber\\
\partial_a g^{b\bar c} \nabla_bW \overline{\nabla_c\nabla_dW} &=& \frac{-4it_2(p^0)^2\hat{D}g_5}{3}\left(\bar{g_1}\hat{D}_{ad}+\bar{g_2}\frac{\hat{D}_a\hat{D}_d}{\hat{D}}\right)\nonumber\\
\partial_{\bar d} g^{b\bar c} \overline{\nabla_cW} \nabla_a\nabla_dW &=&\frac{4it_2(p^0)^2\hat{D}\bar{g_5}}{3}\left({g_1}\hat{D}_{ad}+{g_2}\frac{\hat{D}_a\hat{D}_d}{\hat{D}}\right)\nonumber\\
\partial_a\partial_{\bar d} g^{b\bar c} \nabla_bW \overline{\nabla_cW} &=& \frac{2(p^0)^2g_5\bar{g_5}}{3}\left(3\hat{D}_a\hat{D}_d-2\hat{D}\hat{D}_{ad}\right)  \nonumber
\end{eqnarray} 
Similarly, after simplification,  the individual terms  in $\partial_a\partial_d V$ are given by
\begin{eqnarray}
g^{b\bar c} \nabla_a\nabla_b\nabla_d W \overline{\nabla_cW} &=&\frac{4{t_2}^2(p^0)^2\bar{g_5}\hat{D}}{3}\left[\hat{D}_{ad}\left(6+\frac{15iW}{4p^0\hat{D}{t_2}^3}+\frac{9it}{t_2}+\frac{9t^2}{2{t_2}^2}-\frac{3L}{2{t_2}^2}\right)\right.\nonumber\\
&+&\left.\frac{\hat{D}_a\hat{D}_d}{\hat{D}}\left(\frac{3i(g_1+g_2)}{2t_2}-\frac{27iW}{4p^0\hat{D}{t_2}^3}-\frac{3g_6}{{t_2}^2}-\frac{9t^2}{{t_2}^2}\right.\right.\nonumber\\&+&\left.\left.\frac{9it}{t_2}+\frac{3L}{{t_2}^2}\right)\right]\nonumber
\end{eqnarray}
\begin{eqnarray}
\partial_ag^{b\bar c} \nabla_b\nabla_d W \overline{\nabla_cW} &= &\frac{-4it_2(p^0)^2\hat{D}\bar{g_5}}{3}\left({g_1}\hat{D}_{ad}+{g_2}\frac{\hat{D}_a\hat{D}_d}{\hat{D}}\right)\nonumber\\
3 \nabla_a\nabla_dW\overline{W} &=& 3\overline{W}p^0\left({g_1}\hat{D}_{ad}+{g_2}\frac{\hat{D}_a\hat{D}_d}{\hat{D}}\right)\nonumber\\
- g^{b\bar c} \partial_a g_{d\bar c} \nabla_bW\overline{W} &=&\frac{-ig_5p^0\overline{W}}{t_2}\left(\frac{3\hat{D}_a\hat{D}_d}{\hat{D}}-2\hat{D}_{ad}\right)\nonumber\\
\partial_a\partial_dg^{b\bar c}\nabla_bW\overline{\nabla_cW} &=&\frac{2(p^0)^2|g_5|^2}{3}\left(2\hat{D}\hat{D}_{ad}-3\hat{D}_a\hat{D}_d\right) \nonumber
\end{eqnarray}

We will now substitute the above expressions in eq.(\ref{atwoderiv}) for the two derivative terms of the potential, 
add them up and simplify. For easy reading, we  will 
define the function $\hat F_3(x)$ as 
\begin{equation}
\hat F_3(x) = \sqrt{1 - 24 x (\hat F_1(x))^2} \ ,
\end{equation}
and introduce the parameter $u = \hat F_3(\hat D q_0^2/(\hat\alpha_0^3(p^0)^2)$. 
We find
\begin{eqnarray}
e^{-K_0}\partial_a\partial_dV&=&(p^0\hat\alpha_0)^2\hat{D}_{ad}(3-u)(3+u)^2
\frac{ u^2-2u-1+(u-1) \sqrt{u^2-2 u-3}}{3 \hat{D} (u-1)^2} \cr
e^{-K_0}\partial_a\partial_{\bar{d}}V&=&\frac{2(p^0)^2 (u-3) (u+3)^2 \alpha ^2}{3 \hat{D}^2 (u-1)^2}\left(\hat{D}\hat{D}_{ad}-3\hat{D}_a\hat{D}_d\right)\
  \end{eqnarray}
  Here $K_0$ is the value of the K\"ahler potential at the critical point.  
  
    We can express the mass
matrix in the form
\begin{eqnarray}
  \label{massmatrixappx}
M=E \left(3\frac{\hat{D}_a \hat{D}_d}{  \hat{D}}-\hat{D}_{ad}\right) \otimes {\bf I} +  \hat{D}_{ab} \otimes (A \sigma^3 - B  \sigma^1),
  \end{eqnarray}  
where the coefficients $E, A$ and $B$ are given by 
\begin{eqnarray}
E &=&e^{K_0}\left(\frac{4{\hat{\alpha}_0}^2(p^0)^2(3-u)(u+3)^2}{3\hat{D}(u-1)^2}\right) \cr
A &=& e^{K_0}\left(\frac{2{\hat{\alpha}_0}^2(p^0)^2(3-u)(u+3)^2}{3\hat{D}(u-1)^2}(u^2-2u-1)\right) \cr
B &=&  e^{K_0}\left(\frac{2{\hat{\alpha}_0}^2(p^0)^2(3-u)(u+3)^2}{3\hat{D}(u-1)}\sqrt{(u+1)(3-u)})\right) \ .
\end{eqnarray}


\end{document}